\documentstyle[a4,12pt]{article}
\renewcommand{\baselinestretch}{1.18}
\def\thefootnote{\fnsymbol{footnote}}

\begin{document}
\parskip=5pt plus 1pt minus 1pt

\begin{flushright}
{\bf DPNU-97-23}\\
{April 1997}
\end{flushright}

\vspace{0.2cm}

\begin{center}
{\Large\bf Towards Isolation of New Physics in $B^0_s$-$\bar{B}^0_s$ Mixing}
\end{center}

\vspace{0.3cm}

\begin{center}
\large\bf Zhi-zhong Xing \footnote{Electronic address: xing@eken.phys.nagoya-u.ac.jp}
\end{center}

\begin{center}
{\it Department of Physics, Nagoya University, Chikusa-ku, Nagoya 464-01, Japan}
\end{center}

\vspace{4cm}

\begin{abstract}
In many extensions of the standard model, new physics is possible to contribute 
significantly to $B^0_s$-$\bar{B}^0_s$ mixing. We show that the effect of new
physics, both its phase information and its magnitude, can be isolated from
measurements of $CP$ asymmetries in the semileptonic $B_s$ transitions and in
some nonleptonic $B_s$ decays into hadronic $CP$ eigenstates. We also find that
the rates of $CP$-forbidden decays at the $\Upsilon (5S)$ resonance are not
suppressed by the large $B_s$ mass difference, thus they can be used to extract
the $CP$-violating phase of new physics in either $B^0_s$-$\bar{B}^0_s$ or 
$K^0$-$\bar{K}^0$ mixing.
\end{abstract}

\newpage

\section{Introduction}

Recently some experimental efforts have been made to measure the $B^0_s$-$\bar{B}^0_s$
mixing parameter $x_s$ (corresponding to the mass difference between two $B_s$ 
mass eigenstates), which is anticipated to be very large
in the standard model (SM) \cite{CERN}. This stimulates some theoretical or phenomenological
interest for a more delicate study of the problem of $B^0_s$-$\bar{B}^0_s$ mixing and
$CP$ violation (see, e.g., Refs. \cite{Dunietz,Beneke,Dighe}). In particular, 
the other $B^0_s$-$\bar{B}^0_s$ mixing parameter $y_s$ (corresponding to the width 
difference between two $B_s$ mass eigenstates) is predicted to be about 0.1 within the SM, a 
value making the measurement of untagged $B_s$ data samples meaningful in the
future experiments \cite{Dunietz,Grossman}.

The $B^0_s$-$\bar{B}^0_s$ mixing system is a good place for the 
exploration of new physics (NP) beyond the SM \cite{Review}. For instance, observation of a
sufficiently small $x_s$, an asymmetry between the semileptonic 
$B^0_s$ and $\bar{B}^0_s$ decays, or a significant $CP$-violating signal in
$B^0_s$ vs $\bar{B}^0_s \rightarrow \psi \phi$ modes should imply the existence of
NP in $B^0_s$-$\bar{B}^0_s$ mixing. In most extensitions of the SM, 
NP is expected to contribute only to the mass difference of $B_s$ mesons \cite{Nir}.
Any NP does not significantly affect the direct decays of $B_s$ mesons via the
tree-level $W$-mediated channles, thus the total decay width $\Gamma$ remains to
amount to its SM value as a good approximation. However, the effect of NP can appear
in both $x_s$ and $y_s$, and it may also give rise to a significant $CP$-violating
phase, which is primarily vanishing in the SM.
Consequently it is possible to isolate NP in $B^0_s$-$\bar{B}^0_s$ mixing through
the measurement of $x_s$, $y_s$ as well as $CP$ violation in both semileptonic
and nonleptonic $B_s$ transitions.

Up to now, the possibility to extract the {\it phase} of NP from $CP$ asymmetries in
some $B_s$ decays has been considered (see, e.g., Refs. \cite{Grossman,Cohen}). 
In contrast, little attention has been paid
to determining the {\it magnitude} of NP in $B^0_s$-$\bar{B}^0_s$ mixing, a crucial 
quantity for one to get at the specific non-standard electroweak model.
In this work we shall discuss a few instructive approaches towards the isolation
of NP in $B^0_s$-$\bar{B}^0_s$ mixing, both its phase information and its magnitude.

The remaining parts of this paper are organized as follows: 
in section 2 we describe a parametrization of NP 
in different $B^0_s$-$\bar{B}^0_s$ mixing quantities. The NP effects on $CP$ asymmetries
in semileptonic and nonleptonic decays of $B_s$ mesons are discussed in sections 
3 and 4, respectively. Section 5 is devoted to extracting the phase information of
NP from the $CP$-forbidden $B^0_s\bar{B}^0_s$ transitions. Finally some concluding
remarks are given in section 6.

\section{NP in $B^0_s$-$\bar{B}^0_s$ mixing}

In the assumption of $CPT$ invariance, the mass eigenstates of $B^0_s$ and
$\bar{B}^0_s$ mesons can be written as
\begin{eqnarray}
| B_1 \rangle & = & p |B^0_s\rangle ~ + ~ q |\bar{B}^0_s\rangle \; , \nonumber \\
| B_2 \rangle & = & p |B^0_s\rangle ~ - ~ q |\bar{B}^0_s\rangle \; , 
%		(1)
\end{eqnarray}
where $p$ and $q$ are complex mixing parameters. In terms of off-diagonal
elements of the $2\times 2$ $B^0_s$-$\bar{B}^0_s$ mixing Hamiltonian
${\bf M} - {\rm i} {\bf \Gamma}/2$, we express the ratio $q/p$ as 
\begin{equation}
\frac{q}{p} \; = \; \sqrt{\frac{M^*_{12} ~ - ~ {\rm i} \Gamma^*_{12}/2}
{M_{12} ~ - ~ {\rm i} \Gamma_{12}/2}} \; .
%		(2)
\end{equation}
The mass and width differences of $B_1$ and $B_2$ are defined by
$\Delta m \equiv |m_1 - m_2|$ and $\Delta \Gamma \equiv | \Gamma_1 - \Gamma_2|$,
respectively. Furthermore, we define $m \equiv (m_1 + m_2)/2$ and
$\Gamma \equiv (\Gamma_1 + \Gamma_2)/2$. It is also convenient
to use the following dimensionless parameters to describe $B^0_s$-$\bar{B}^0_s$
mixing:
\begin{equation}
x_s \; \equiv \; \frac{\Delta m}{\Gamma} \; , ~~~~~~~~
y_s \; \equiv \; \frac{\Delta \Gamma}{ 2\Gamma} \; .
%		(3)
\end{equation}
The present experimental lower bound for $x_s$ is $x_s \geq 15$ \cite{CERN}, implying
$\Delta m \gg \Delta \Gamma$ or $|M_{12}| \gg |\Gamma_{12}|$ holds. 
Consequently one gets 
\begin{equation}
\Delta m \; = \; 2 |M_{12}| 
%		(4)
\end{equation}
to a high degree of accuracy. Then the width difference $\Delta \Gamma$ reads \cite{Grossman}
\begin{equation}
\Delta \Gamma \; = \; \frac{4 | {\rm Re} (M_{12} \Gamma^*_{12} )|}
{\Delta m} \; = \; 2 |\Gamma_{12}| \cdot |\cos\phi_m| \; ,
%		(5)
\end{equation}
where $\phi_m \equiv \arg (-M_{12}/\Gamma_{12})$. In addition, calculations
based on the SM yields $y_s \approx 0.08$ \cite{Beneke}. This value will always be reduced 
if $B^0_s$-$\bar{B}^0_s$ mixing receives $CP$-violating contributions 
from NP (i.e., $\phi_m \neq 0$). Thus $x_s/y_s > 10^2$ is true both within and
beyond the SM.

Now we assume that NP contributes to $\Delta m$ through $M_{12}$, which can be
parametrized as
\begin{equation}
M_{12} \; =\; M^{\rm SM}_{12} ~ + ~ M^{\rm NP}_{12} \; =\; 
M^{\rm SM}_{12} \left ( 1 + z e^{{\rm i}\theta} \right ) \;
%		(6)
\end{equation}
with $z\equiv |M^{\rm NP}_{12}/M^{\rm SM}_{12}|$ and $\theta \equiv \arg 
(M^{\rm NP}_{12}/M^{\rm SM}_{12})$. The effect of NP on $\Gamma_{12}$ 
is anticipated to be negligibly small, hence $\Gamma_{12} = \Gamma^{\rm SM}_{12}$
and $\Gamma = \Gamma^{\rm SM}$ hold as a good approximation \cite{Review}. Since
$M^{\rm SM}_{12}$ and $\Gamma^{\rm SM}_{12}$ are dominantly proportional to
the Cabibbo-Kobayashi-Maskawa 
matrix elements $V_{tb}V^*_{ts}$ and $V_{cb}V^*_{cs}$, respectively,
$\phi_m =0$ turns out to be an excellent approximation in the Wolfenstein
phase convention. The presence of NP in $M_{12}$ modifies the SM value of
$\phi_m$ in the following manner:
\begin{equation}
\phi_m \; =\; \arg \left (1 + z e^{{\rm i}\theta} \right ) \; =\; 
\arctan \left [ \frac{z \sin\theta}{1 + z\cos\theta} \right ] \; .
%		(7)
\end{equation}
The $B^0_s$-$\bar{B}^0_s$ mixing observables $x_s$ and $y_s$, in the
existence of NP, read
\begin{equation}
x_s \; = \; x^{\rm SM}_s \sqrt{ 1 + z^2 + 2 z \cos\theta} \; 
%		(8)
\end{equation}
and
\begin{equation}
y_s \; = \; y^{\rm SM}_s \sqrt{ 1 - \frac{z^2 \sin^2\theta}{1 + z^2 + 2z \cos\theta}} \; ,
%		(9)
\end{equation}
where $x^{\rm SM}_s$ and $y^{\rm SM}_s$ are (in principle) calculable from the SM.
Finally, the ratio $q/p$ can be expressed as
\begin{eqnarray}
\frac{q}{p} & = & \displaystyle \sqrt{\frac{ \left ( 1+ z e^{-{\rm i}\theta}
\right ) ~ + ~ {\rm i} y^{\rm SM}_s /x^{\rm SM}_s}
{ \left ( 1+ z e^{+{\rm i}\theta} \right ) ~ + ~ {\rm i} y^{\rm SM}_s /
x^{\rm SM}_s}} \; \nonumber \\
& \approx & e^{-{\rm i}2\phi_m} \left [ 1 ~ - ~ \frac{z\sin \theta}{1 + z^2
+ 2 z \cos\theta} \cdot \frac{y^{\rm SM}_s}{x^{\rm SM}_s} \right ] \; 
%		(10)
\end{eqnarray}
in the next-to-leading order approximation. Note that $\phi_m$ is indeed
a function of $z$ and $\theta$, as given in Eq. (7). Clearly $\arg (q/p)
= -2\phi_m$ is a good approximation when we discuss $CP$ violation in nonleptonic 
$B_s$ decays. The small correction term in $q/p$, which is proportional to
$z\sin\theta$ and $y^{\rm SM}_s/x^{\rm SM}_s$, is crucial for an asymmetry
between the semileptonic $B^0_s$ and $\bar{B}^0_s$ transitions (see section 3).

One can see, from Eq. (9), that $y_s \leq y^{\rm SM}_s$ is always true. In 
contrast, $x_s \leq x^{\rm SM}_s$ takes place only if $\theta \in [-\pi, -\pi/2]$
or $[\pi/2, \pi]$, corresponding to the destructive interference between
$M^{\rm NP}_{12}$ and $M^{\rm SM}_{12}$. Provided $z\gg 1$ (i.e., NP
dominates $B^0_s$-$\bar{B}^0_s$ mixing), we obtain $x_s \approx z x^{\rm SM}_s$
and $y_s \approx y^{\rm SM}_s \cos\theta$. In this case, $q/p = e^{-{\rm i}2\phi_m}$
holds to a higher degree of accuracy.

\section{NP effect in semileptonic $B_s$ decays}

Now we examine the NP effect on $CP$ violation in semileptonic transitions of
$B_s$ mesons. Assuming both the $\Delta Q = \Delta B$ rule and $CPT$ invariance,
one is able to get the time-integrated wrong-sign events of semileptonic $B_s$
decays as follows \cite{Xing96PR,Xing97}:
\begin{eqnarray}
N^-(B^0_s) & = & n_0 \left [ \frac{1}{1-y^2_s} ~ - ~ \frac{1}{1+x^2_s} \right ]
\left | \frac{q}{p} \right |^2 \; , \nonumber \\
N^+(\bar{B}^0_s) & = & n_0 \left [ \frac{1}{1-y^2_s} ~ - ~ \frac{1}{1+x^2_s} \right ]
\left | \frac{p}{q} \right |^2 \; , 
%		(11)
\end{eqnarray}
where $n_0$ is a normalization factor proportional to the rate of the right-sign
semileptonic $B_s$ decay. The $CP$ asymmetry between the above two processes
turns out to be
\begin{equation}
{\cal A}_{\rm SL} \; \equiv \; \frac{N^+(\bar{B}^0_s) ~ - ~ N^-(B^0_s)}
{N^+(\bar{B}^0_s) ~ + ~ N^-(B^0_s)} \; = \;
\frac{|p|^4 ~ - ~ |q|^4}{|p|^4 ~ + ~ |q|^4} \; ,
%		(12)
\end{equation}
apparently irrelevant to the mixing parameters $x_s$ and $y_s$.
 
Such an asymmetry can also be obtained from the same-sign dilepton events of coherent
$B^0_s\bar{B}^0_s$ decays at the $\Upsilon (5S)$ resonance, whose rates are
given by \cite{Xing97,Xing96PL}
\begin{eqnarray}
N^{--}_C (B^0_s\bar{B}^0_s) & = & n^{~}_C \left [ \frac{1+Cy^2_s}{(1-y^2_s)^2}
~ - ~ \frac{1-Cx^2_s}{(1+x^2_s)^2} \right ] \left | \frac{q}{p} \right |^2 \; ,
\nonumber \\
N^{++}_C (B^0_s\bar{B}^0_s) & = & n^{~}_C \left [ \frac{1+Cy^2_s}{(1-y^2_s)^2}
~ - ~ \frac{1-Cx^2_s}{(1+x^2_s)^2} \right ] \left | \frac{p}{q} \right |^2 \; ,
%		(13)
\end{eqnarray}
in which $C$ ($=\pm 1$) is the charge-conjugation parity of the $B^0_s\bar{B}^0_s$
pair, and $n_C$ denotes the normalization factor proportional to the rates of
right-sign semileptonic decays of $B^0_s$ and $\bar{B}^0_s$ mesons. Obviously,
we arrive at the same $CP$ asymmetry ${\cal A}_{\rm SL}$:
\begin{equation}
\frac{N^{++}_C(B^0_s\bar{B}^0_s) ~ - ~ 
N^{--}_C(B^0_s\bar{B}^0_s)}{N^{++}_C(B^0_s\bar{B}^0_s) ~ + ~ N^{--}_C(B^0_s\bar{B}^0_s)} 
\; = \; \frac{|p|^4 ~ - ~ |q|^4}{|p|^4 ~ + ~ |q|^4} \; ,
%		(14)
\end{equation}
for both $C=-1$ and $C=+1$ cases.

Taking $|M_{12}|\gg |\Gamma_{12}|$ into account, one can obtain ${\cal A}_{\rm SL}
\approx {\rm Im} (\Gamma_{12}/M_{12})$ as a good approximation. 
By use of Eq. (10), we find that ${\cal A}_{\rm SL}$ explicitly reads
\begin{equation}
{\cal A}_{\rm SL} \; \approx \; \frac{2 z \sin\theta}{1+z^2 +2 z\cos\theta}
\cdot \frac{y^{\rm SM}_s}{x^{\rm SM}_s} \; .
%		(15)
\end{equation}
Within the SM (i.e., $z=0$ and $\theta =0$), ${\cal A}_{\rm SL}$ is vanishingly
small. In view of the fact $y^{\rm SM}_s/x^{\rm SM}_s \sim O(10^{-2})$, 
we believe that the presence of NP in $B^0_s$-$\bar{B}^0_s$ mixing could
enhance the asymmetry ${\cal A}_{\rm SL}$ to the level of $10^{-3} - 10^{-2}$.
For the purpose of illustration, we plot changes of ${\cal A}_{\rm SL}$ with
the $CP$-violating phase $\theta$ by assuming $y^{\rm SM}_s/x^{\rm SM}_s =
0.05$ and $z=0.5$ (or $z=5$) in Fig. 1. It is clear that ${\cal A}_{\rm SL}$ may
reach the percent level for favorable values of $\theta$ and $z$. Thus the
measurement of ${\cal A}_{\rm SL}$ at the forthcoming $B$-meson factories can provide 
some useful information or constraints on NP in $B^0_s$-$\bar{B}^0_s$ mixing.
%%%%%%%%%%%%%%%%%%%%%%%%%%%%%%%%%%%%%
\begin{figure}
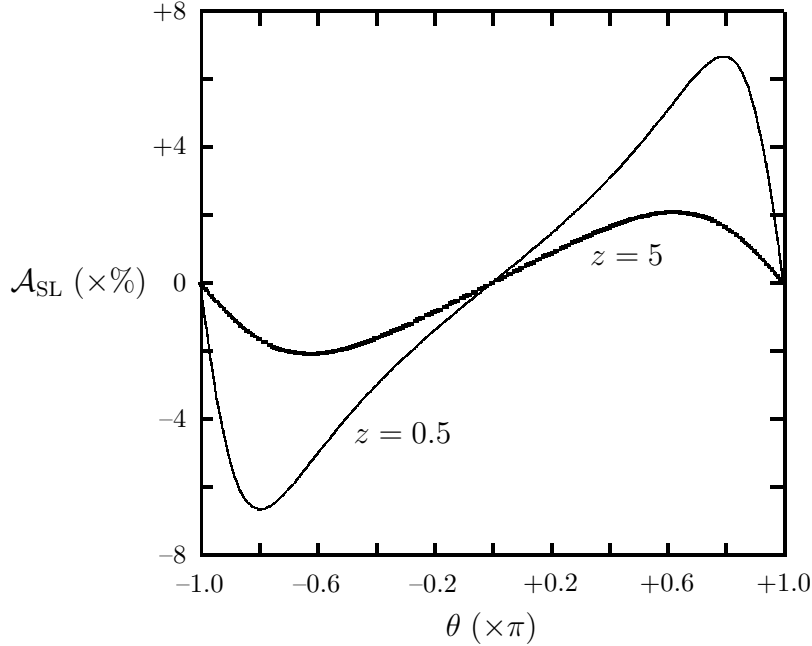

% GNUPLOT: LaTeX picture
\setlength{\unitlength}{0.240900pt}
\ifx\plotpoint\undefined\newsavebox{\plotpoint}\fi
\sbox{\plotpoint}{\rule[-0.500pt]{1.000pt}{1.000pt}}%

\vspace{0.4cm}
\caption{Illustrative plot for changes of the semileptonic $CP$
asymmetry ${\cal A}_{\rm SL}$ with $\theta$, where
$y^{\rm SM}_s/x^{\rm SM}_s = 0.05$ and $z=0.5$ or $z=5$ have been taken.}
\end{figure}
%%%%%%%%%%%%%%%%%%%%%%%%%%%%%%%%%%%

Finally it is worth emphasizing that reliable calculations of $y^{\rm SM}_s/x^{\rm SM}_s$
are crucial for probing any NP effect through ${\cal A}_{\rm SL}$. A detailed analysis of
$y^{\rm SM}_s/x^{\rm SM}_s$ with updated data can be found in Ref. \cite{Xing98}.

\section{NP effect in nonleptonic $B_s$ decays}

Let us turn attention to the NP effect on $CP$ asymmetries in some nonleptonic
$B_s$ decays into $CP$ eigenstates, such as $D^+_s D^-_s$, $D^{*+}_sD^-_s \oplus 
D^+_s D^{*-}_s$, $(\psi \phi)_+$ (i.e., the $CP$-even state of $\psi \phi$) \cite{Dighe},
$\psi K_S$ and $\psi K_L$. These transitions 
can be used to extract the $CP$-violating phase $\phi_m$, in most cases, without
ambiguities from hadronic matrix elements. They are also favorable in view of the
nearest experimental detectability for weak decays of $B_s$ mesons at $e^+e^-$
colliders and (or) hadron machines. Obviously, $q/p = e^{-{\rm i}2\phi_m}$ should 
be an excellent approximation. 

Due to $B^0_s$-$\bar{B}^0_s$ mixing, the time-dependent rates of $B_s$ mesons 
decaying into a $CP$ eigenstate $f$ can be written as \cite{Xing97}
\begin{eqnarray}
{\cal R} [ B^0_{s, \rm phys} (t) \rightarrow f] & = & n^{~}_f e^{-\Gamma t}
\left [ \frac{1+|\lambda_f|^2}{2} \cosh (y_s \Gamma t) ~ - ~
{\rm Re} \lambda_f \sinh (y_s \Gamma t) \right . \nonumber \\
&  & ~~~~~~~ \left . + ~ \frac{1-|\lambda_f|^2}{2} \cos (x_s \Gamma t) ~ - ~
{\rm Im}\lambda_f \sin (x_s \Gamma t) \right ] \; , \nonumber \\
{\cal R} [ \bar{B}^0_{s, \rm phys} (t) \rightarrow f] & = & n^{~}_f e^{-\Gamma t}
\left [ \frac{1+|\lambda_f|^2}{2} \cosh (y_s \Gamma t) ~ - ~
{\rm Re} \lambda_f \sinh (y_s \Gamma t) \right . \nonumber \\
&  & ~~~~~~~ \left . - ~ \frac{1-|\lambda_f|^2}{2} \cos (x_s \Gamma t) ~ + ~
{\rm Im}\lambda_f \sin (x_s \Gamma t) \right ] \; ,
%		(16)
\end{eqnarray}
where $n^{~}_f$ denotes the normalization factor proportional to 
$|\langle f|{\cal H}_{\rm eff}|B^0_s\rangle|^2$, and $\lambda_f$ 
stands for a rephasing-invariant quantity defined by
\begin{equation}
\lambda_f \; \equiv \; e^{-{\rm i}2\phi_m} \frac{\langle f|{\cal H}_{\rm eff}|
\bar{B}^0_s\rangle}{\langle f|{\cal H}_{\rm eff}|B^0_s\rangle} \; .
%		(17)
\end{equation}
We see that the $x_s$-induced oscillations, anticipated to be rapid in the SM, 
cancel in the rate of the flavor-untagged data samples:
\begin{eqnarray}
{\cal S}(t) & \equiv & {\cal R}[B^0_{s,\rm phys} (t)\rightarrow f]
~ + ~ {\cal R}[\bar{B}^0_{s,\rm phys} (t) \rightarrow f] \; , \nonumber \\
& = & n^{~}_f e^{-\Gamma t} \left [ \left (1+|\lambda_f|^2 \right ) \cosh (y_s \Gamma t)
~ - ~ 2  {\rm Re}\lambda_f \sinh (y_s \Gamma t) \right ] \; .
%		(18)
\end{eqnarray}
In contrast, the $y_s$-induced oscillations disappear in the rate asymmetry 
between the flavor-tagged data sample:
\begin{eqnarray}
{\cal A}(t) & \equiv & {\cal R}[B^0_{s,\rm phys} (t)\rightarrow f]
~ - ~ {\cal R}[\bar{B}^0_{s,\rm phys} (t) \rightarrow f] \; , \nonumber \\
& = & n^{~}_f e^{-\Gamma t} \left [ \left (1-|\lambda_f|^2 \right ) \cos (x_s \Gamma t)
~ - ~ 2  {\rm Im}\lambda_f \sin (x_s \Gamma t) \right ] \; .
%		(19)
\end{eqnarray}
Therefore, measurements of ${\cal S}(t)$ and (or) ${\cal A}(t)$
allow one to determine ${\rm Re}\lambda_f$ versus $(1+|\lambda_f|^2)$ and (or) 
${\rm Im}\lambda_f$ versus $(1-|\lambda_f|^2)$ in most cases, except the special situation
where $y_s$ is vanishingly small (e.g., $y_s\leq 10^{-2}$) due to NP and
$x_s$ is considerably large (e.g., $x_s > 50$). 

Within the SM, $\lambda_{D^+_s D^-_s} \approx 1$, $\lambda_{(\psi \phi)_+} \approx 1$,
$\lambda_{\psi K_S} \approx -1$ and $\lambda_{\psi K_L} \approx 1$ hold as a good
approximation. For these decay modes, the strong or electroweak penguin contributions
(dominated by the top quark in the loop) are significantly suppressed in magnitude 
and do not give rise to additional $CP$-violating phases. Beyond the SM, the NP
appearing in $B^0_s$-$\bar{B}^0_s$ mixing is likely to simultaneously manifest itself 
in the loop-induced penguin amplitudes. Generally we do not expect that the NP-enhanced
penguin contributions can be significant enough in those $B_s$ transitions mentioned above,
and a test of this assumption is possible by detecting the deviation of $|\lambda_f|$
from unity in ${\cal A}(t)$. Here we assume $|\lambda_f|\approx 1$
to hold even in the presence of NP, for $B_s\rightarrow D^+_sD^-_s$, $(\psi \phi)_+$,
$\psi K_S$ and $\psi K_L$. Then we obtain
\begin{eqnarray}
{\rm Re} \lambda_{D^+_s D^-_s} & \approx & {\rm Re}\lambda_{(\psi \phi)_+} \; \approx \;
- {\rm Re}\lambda_{\psi K_S} \; \approx \; {\rm Re} \lambda_{\psi K_L} \; \approx \;
\cos(2\phi_m) \; , \nonumber \\
{\rm Im} \lambda_{D^+_s D^-_s} & \approx & {\rm Im}\lambda_{(\psi \phi)_+} \; \approx \;
- {\rm Im}\lambda_{\psi K_S} \; \approx \; {\rm Im} \lambda_{\psi K_L} \; \approx \;
- \sin (2\phi_m) \; ,
%		(20)
\end{eqnarray}
where
\begin{equation}
\cos (2\phi_m) \; = \; \frac{1 +2z \cos\theta + z^2 \cos (2\theta)}{1 + z^2 + 2z\cos \theta} \; ,
~~~~~~ \sin (2\phi_m) \; = \; \frac{2 z\sin\theta \left (1 + z\cos\theta \right )}{1 +z^2 +
2z \cos\theta} \; .
%		(21)
\end{equation}
Through this way, we can determine the phase information of NP in $B^0_s$-$\bar{B}^0_s$
mixing. To illustrate the correlation between two observables, e.g., 
${\rm Re}\lambda_{(\psi \phi)_+}$
and ${\rm Im}\lambda_{(\psi \phi)_+}$, we plot them in Fig. 2 by taking $\theta \in [-\pi, +\pi]$
and $z=0.5$ or $z=5$. Clearly the two-fold ambiguity in extracting $\phi_m$ from
${\rm Re}\lambda_{(\psi \phi)_+} \approx \cos (2\phi_m)$ can be removed if 
${\rm Im}\lambda_{(\psi \phi)_+} \approx - \sin (2\phi_m)$
is also measured. 
%%%%%%%%%%%%%%%%%%%%%%%%%%%%%%%%%%%%%
\begin{figure}
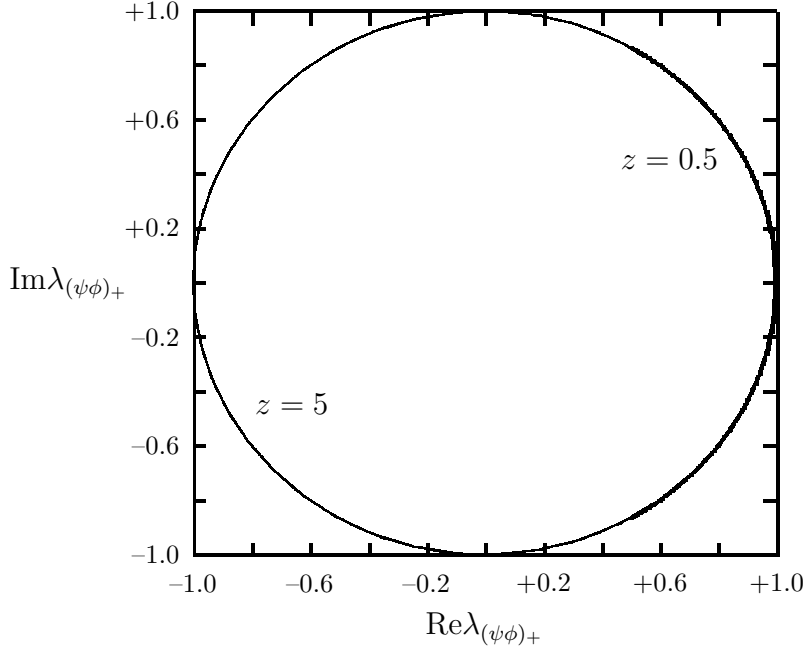

% GNUPLOT: LaTeX picture
\setlength{\unitlength}{0.240900pt}
\ifx\plotpoint\undefined\newsavebox{\plotpoint}\fi
\sbox{\plotpoint}{\rule[-0.500pt]{1.000pt}{1.000pt}}%

\vspace{0.4cm}
\caption{Illustrative plot for the correlation between two observables
${\rm Re}\lambda_{(\psi \phi)_+}$ and ${\rm Im}\lambda_{(\psi \phi)_+}$,
where $\theta \in [-\pi, +\pi]$ and $z=0.5$ (dark solid curve) or
$z=5$ (solid circle) have been taken.}
\end{figure}
%%%%%%%%%%%%%%%%%%%%%%%%%%%%%%%%%%%%%%%%%

It is obvious that a determination of $\phi_m$ itself cannot
isolate the magnitude ($z$) and phase ($\theta$) of NP effects in $B^0_s$-$\bar{B}^0_s$ mixing.
For this reason, combining the measurements of ${\cal A}_{\rm SL}$, ${\cal S}(t)$
and (or) ${\cal A}(t)$ is useful to determine or constrain $z$ and $\theta$
separately. 

\section{NP effect in $CP$-forbidden $B^0_s\bar{B}^0_s$ decays}

If the expected large size of $x^{\rm SM}_s$ cannot be remarkably reduced by the contribution
from NP in $B^0_s$-$\bar{B}^0_s$ mixing, then a study of $CP$-forbidden $B^0_s\bar{B}^0_s$
decays at the $\Upsilon (5S)$ resonance will become interesting. The straightforward reason 
is that the rates of such $CP$-forbidden transitions are not suppressed by the largeness of
$x_s$, as we shall show subsequently. The $CP$-violating signal may be established directly
from the observation of a $CP$-forbidden decay mode itself other than the decay rate
asymmetry, thus neither flavor tagging nor time-dependent measurements are necessary in
practical experiments \cite{Xing96PR}. 
For a $B$-meson factory running at the $\Upsilon (5S)$ resonance, 
one will have a chance to detect the typical $CP$-forbidden channels
like $(B^0_s\bar{B}^0_s)_{C=+1}\rightarrow (\psi K_S) (\psi K_L)$ and
$(B^0_s\bar{B}^0_s)_{C=-1}\rightarrow (\psi K_S) (\psi K_S)$ or $(\psi K_L) (\psi K_L)$,
where $CP$ parities of the initial and final states are opposite.
For simplicity and illustration, we concentrate only on the coherent decays of 
$(B^0_s\bar{B}^0_s)_{C=-1}$ events later on.

The time-independent rate of a $(B^0_s\bar{B}^0_s)_{C=-1}$ pair decaying into the 
final state $(f_1 f_2)$ can be given by \cite{Xing96PR,Xing97}
\begin{eqnarray}
{\cal R}(f_1, f_2) & = & n^{~}_{12} \left \{ \frac{x^2_s}{1+x^2_s} \left [
1 + |\lambda_{f_1}|^2 |\lambda_{f_2}|^2 - 2 {\rm Re} \left (\lambda_{f_1} \lambda_{f_2}
\right ) \right ] \; \right . \nonumber \\
&  & \left . ~~~ + ~ \frac{2 +x^2_s}{1+x^2_s} \left [ |\lambda_{f_1}|^2 + |\lambda_{f_2}|^2
- 2{\rm Re} \left (\lambda_{f_1} \lambda_{f_2}^* \right ) \right ] \right \} \; ,
%		(22)
\end{eqnarray}
where $n^{~}_{12}$ is a normalization factor proportional to the product of 
${\cal R}(B^0_s \rightarrow f_1)$ and ${\cal R}(B^0_s\rightarrow f_2)$; and 
$\lambda_{f_i}$ ($i=1$ or $2$) is defined like $\lambda_f$ in Eq. (17). In obtaining
the above formula, we have made a safe approximation: $1/(1-y^2_s) \approx 1$
due to $y_s \leq y^{\rm SM}_s \sim 0.1$. For those $B_s$ decays into hadronic $CP$
eigenstates via quark subprocesses $b\rightarrow (c\bar{c})s$ and $b\rightarrow
(c\bar{c})d$, $|\lambda_{f_1}| \approx |\lambda_{f_2}|\approx 1$ is expected to
hold to a good degree of accuracy. With the help of ${\cal R}(B^0_s\rightarrow \psi K_S)
\approx {\cal R}(B^0_s\rightarrow \psi K_L)$, one can calculate the following ratios
of two joint decay rates:
\begin{equation}
\xi_{1K} \; \equiv \; \frac{{\cal R}(D^+_sD^-_s, \psi K_L)}{{\cal R}
(D^+_sD^-_s, \psi K_S)} \; , ~~~~~~~~
\xi_{2K} \; \equiv \; \frac{{\cal R}(\psi K_S, \psi K_S)}{{\cal R}
(\psi K_S, \psi K_L)} \; .
%		(23)
\end{equation}
The results explicitly read
\begin{eqnarray}
\xi_{1K} & = & \frac{x^2_s \left [ 1 - {\rm Re}(\lambda_{D^+_sD^-_s} \lambda_{\psi K_L})
\right ] + (2 +x^2_s) \left [ 1- {\rm Re}(\lambda_{D^+_sD^-_s} \lambda^*_{\psi K_L})
\right ]}
{x^2_s \left [ 1 - {\rm Re}(\lambda_{D^+_sD^-_s} \lambda_{\psi K_S})
\right ] + (2 +x^2_s) \left [ 1- {\rm Re}(\lambda_{D^+_sD^-_s} \lambda^*_{\psi K_S})
\right ]} \; , \nonumber \\
& = & \frac{x^2_s \sin^2 (2\phi_m - \phi_K) + (2+x^2_s) \sin^2\phi_K}
{x^2_s \cos^2(2\phi_m -\phi_K) + (2+x^2_s) \cos^2\phi_K} \; ;
%		(24)
\end{eqnarray}
and
\begin{eqnarray}
\xi_{2K} & = & \frac{x^2_s \left [ 1 - {\rm Re}(\lambda^2_{\psi K_S} )
\right ] + (2 +x^2_s) \left [ 1- {\rm Re}(|\lambda_{\psi K_S}|^2)
\right ]}
{x^2_s \left [ 1 - {\rm Re}(\lambda_{\psi K_S} \lambda_{\psi K_L})
\right ] + (2 +x^2_s) \left [ 1- {\rm Re}(\lambda_{\psi K_S} \lambda^*_{\psi K_L})
\right ]} \; , \nonumber \\
& = & \frac{x^2_s \sin^2 [2(\phi_m - \phi_K)]}
{x^2_s \cos^2[2(\phi_m - \phi_K)] + (2+x^2_s) } \; ,
%		(25)
\end{eqnarray}
where the phase parameter $\phi_K$ comes from $K^0$-$\bar{K}^0$ mixing
in the final states of $B_s \rightarrow \psi K_S$ and $\psi K_L$: 
$2\phi_K \equiv -\arg (q_K/p_K)$. Within the SM, $\phi_K
\approx 0$ is a good approximation. However, the NP appearing in $B^0_s$-$\bar{B}^0_s$
mixing is also likely to manifest itself in $K^0$-$\bar{K}^0$ mixing, leading probably to a
significant $CP$-violating phase $\phi_K$. One can see that the above two
$CP$-violating signals are not suppressed by the largeness of $x_s$. For $x_s \geq 15$ \cite{CERN},
we approximate Eqs. (24) and (25) to
\begin{eqnarray}
\xi_{1K} & \approx & \frac{\sin^2 (2\phi_m - \phi_K) + \sin^2\phi_K}
{\cos^2(2\phi_m -\phi_K) + \cos^2\phi_K} \; , \nonumber \\
\xi_{2K} & \approx & \frac{\sin^2 [2(\phi_m - \phi_K)]}
{\cos^2[2(\phi_m - \phi_K)] + 1} \; .
%		(26)
\end{eqnarray}
Clearly the information on $\phi_m$ and $\phi_K$ is extractable from $\xi_{1K}$
and $\xi_{2K}$. If $\phi_K \approx  0$ can be further assumed, then Eq. (26)
is simplified as 
\begin{equation}
\xi_{1K} \;  \approx \; \xi_{2K} \; \approx  \; \frac{\sin^2(2\phi_m)}{1+\cos^2(2\phi_m)} \; ,
%		(27)
\end{equation}
which is a pure function of $\phi_m$. 

For the purpose of illustration, we assume $\phi_m -\phi_K = \pi/6$ or
$\pi/4$ to plot the change of $\xi_{2K}$ with $x_s \in [1, 30]$ in Fig. 3.
We find that $\xi_{2K}$ becomes close to its maximal value only if 
$x_s \geq 10$, for a definite input of $\phi_m -\phi_K$. Hence 
such a $CP$-violating signal will be of particular experimental interest, 
provided NP introduces a significant phase into $B^0_s$-$\bar{B}^0_s$ (or 
$K^0$-$\bar{K}^0$) mixing.
%%%%%%%%%%%%%%%%%%%%%%%%%%%%%%%%%%%%%%%%%%%
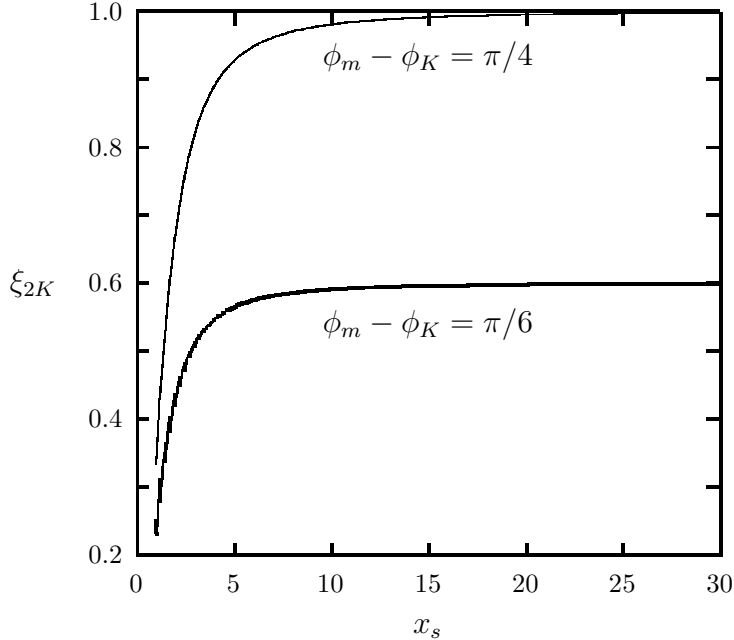
\begin{figure}
% GNUPLOT: LaTeX picture
\setlength{\unitlength}{0.240900pt}
\ifx\plotpoint\undefined\newsavebox{\plotpoint}\fi
\sbox{\plotpoint}{\rule[-0.500pt]{1.000pt}{1.000pt}}%
\begin{picture}(1200,990)(-350,0)
\font\gnuplot=cmr10 at 10pt
\gnuplot
\sbox{\plotpoint}{\rule[-0.500pt]{1.000pt}{1.000pt}}%
\put(220.0,113.0){\rule[-0.500pt]{1.000pt}{205.729pt}}
\put(220.0,113.0){\rule[-0.500pt]{4.818pt}{1.000pt}}
\put(198,113){\makebox(0,0)[r]{0.2}}
\put(1116.0,113.0){\rule[-0.500pt]{4.818pt}{1.000pt}}
\put(220.0,220.0){\rule[-0.500pt]{4.818pt}{1.000pt}}
%\put(198,220){\makebox(0,0)[r]{0.3}}
\put(1116.0,220.0){\rule[-0.500pt]{4.818pt}{1.000pt}}
\put(220.0,327.0){\rule[-0.500pt]{4.818pt}{1.000pt}}
\put(198,327){\makebox(0,0)[r]{0.4}}
\put(1116.0,327.0){\rule[-0.500pt]{4.818pt}{1.000pt}}
\put(220.0,433.0){\rule[-0.500pt]{4.818pt}{1.000pt}}
%\put(198,433){\makebox(0,0)[r]{0.5}}
\put(1116.0,433.0){\rule[-0.500pt]{4.818pt}{1.000pt}}
\put(220.0,540.0){\rule[-0.500pt]{4.818pt}{1.000pt}}
\put(198,540){\makebox(0,0)[r]{0.6}}
\put(1116.0,540.0){\rule[-0.500pt]{4.818pt}{1.000pt}}
\put(220.0,647.0){\rule[-0.500pt]{4.818pt}{1.000pt}}
%\put(198,647){\makebox(0,0)[r]{0.7}}
\put(1116.0,647.0){\rule[-0.500pt]{4.818pt}{1.000pt}}
\put(220.0,753.0){\rule[-0.500pt]{4.818pt}{1.000pt}}
\put(198,753){\makebox(0,0)[r]{0.8}}
\put(1116.0,753.0){\rule[-0.500pt]{4.818pt}{1.000pt}}
\put(220.0,860.0){\rule[-0.500pt]{4.818pt}{1.000pt}}
%\put(198,860){\makebox(0,0)[r]{0.9}}
\put(1116.0,860.0){\rule[-0.500pt]{4.818pt}{1.000pt}}
\put(220.0,967.0){\rule[-0.500pt]{4.818pt}{1.000pt}}
\put(198,967){\makebox(0,0)[r]{1.0}}
\put(1116.0,967.0){\rule[-0.500pt]{4.818pt}{1.000pt}}
\put(220.0,113.0){\rule[-0.500pt]{1.000pt}{4.818pt}}
\put(220,68){\makebox(0,0){0}}
\put(220.0,947.0){\rule[-0.500pt]{1.000pt}{4.818pt}}
\put(373.0,113.0){\rule[-0.500pt]{1.000pt}{4.818pt}}
\put(373,68){\makebox(0,0){5}}
\put(373.0,947.0){\rule[-0.500pt]{1.000pt}{4.818pt}}
\put(525.0,113.0){\rule[-0.500pt]{1.000pt}{4.818pt}}
\put(525,68){\makebox(0,0){10}}
\put(525.0,947.0){\rule[-0.500pt]{1.000pt}{4.818pt}}
\put(678.0,113.0){\rule[-0.500pt]{1.000pt}{4.818pt}}
\put(678,68){\makebox(0,0){15}}
\put(678.0,947.0){\rule[-0.500pt]{1.000pt}{4.818pt}}
\put(831.0,113.0){\rule[-0.500pt]{1.000pt}{4.818pt}}
\put(831,68){\makebox(0,0){20}}
\put(831.0,947.0){\rule[-0.500pt]{1.000pt}{4.818pt}}
\put(983.0,113.0){\rule[-0.500pt]{1.000pt}{4.818pt}}
\put(983,68){\makebox(0,0){25}}
\put(983.0,947.0){\rule[-0.500pt]{1.000pt}{4.818pt}}
\put(1136.0,113.0){\rule[-0.500pt]{1.000pt}{4.818pt}}
\put(1136,68){\makebox(0,0){30}}
\put(1136.0,947.0){\rule[-0.500pt]{1.000pt}{4.818pt}}
\put(220.0,113.0){\rule[-0.500pt]{220.664pt}{1.000pt}}
\put(1136.0,113.0){\rule[-0.500pt]{1.000pt}{205.729pt}}
\put(220.0,967.0){\rule[-0.500pt]{220.664pt}{1.000pt}}
\put(58,540){\makebox(0,0){$\xi_{2K}$}}
\put(678,-3){\makebox(0,0){$x_s$}}
\put(678,895){\makebox(0,0){$\phi_m -\phi_K = \pi/4$}}
\put(678,475){\makebox(0,0){$\phi_m -\phi_K = \pi/6$}}
\put(220.0,113.0){\rule[-0.500pt]{1.000pt}{205.729pt}}
\put(251,146){\usebox{\plotpoint}}
\multiput(252.84,146.00)(0.475,5.490){6}{\rule{0.114pt}{10.250pt}}
\multiput(248.92,146.00)(7.000,48.726){2}{\rule{1.000pt}{5.125pt}}
\multiput(259.83,216.00)(0.481,3.838){8}{\rule{0.116pt}{7.500pt}}
\multiput(255.92,216.00)(8.000,42.433){2}{\rule{1.000pt}{3.750pt}}
\multiput(267.84,274.00)(0.475,3.537){6}{\rule{0.114pt}{6.821pt}}
\multiput(263.92,274.00)(7.000,31.842){2}{\rule{1.000pt}{3.411pt}}
\multiput(274.83,320.00)(0.481,2.395){8}{\rule{0.116pt}{4.875pt}}
\multiput(270.92,320.00)(8.000,26.882){2}{\rule{1.000pt}{2.438pt}}
\multiput(282.83,357.00)(0.481,1.845){8}{\rule{0.116pt}{3.875pt}}
\multiput(278.92,357.00)(8.000,20.957){2}{\rule{1.000pt}{1.938pt}}
\multiput(290.84,386.00)(0.475,1.665){6}{\rule{0.114pt}{3.536pt}}
\multiput(286.92,386.00)(7.000,15.661){2}{\rule{1.000pt}{1.768pt}}
\multiput(297.83,409.00)(0.481,1.158){8}{\rule{0.116pt}{2.625pt}}
\multiput(293.92,409.00)(8.000,13.552){2}{\rule{1.000pt}{1.313pt}}
\multiput(305.83,428.00)(0.481,0.883){8}{\rule{0.116pt}{2.125pt}}
\multiput(301.92,428.00)(8.000,10.589){2}{\rule{1.000pt}{1.063pt}}
\multiput(313.84,443.00)(0.475,0.851){6}{\rule{0.114pt}{2.107pt}}
\multiput(309.92,443.00)(7.000,8.627){2}{\rule{1.000pt}{1.054pt}}
\multiput(320.83,456.00)(0.481,0.539){8}{\rule{0.116pt}{1.500pt}}
\multiput(316.92,456.00)(8.000,6.887){2}{\rule{1.000pt}{0.750pt}}
\multiput(328.83,466.00)(0.481,0.470){8}{\rule{0.116pt}{1.375pt}}
\multiput(324.92,466.00)(8.000,6.146){2}{\rule{1.000pt}{0.688pt}}
\multiput(335.00,476.84)(0.362,0.475){6}{\rule{1.250pt}{0.114pt}}
\multiput(335.00,472.92)(4.406,7.000){2}{\rule{0.625pt}{1.000pt}}
\multiput(342.00,483.84)(0.476,0.462){4}{\rule{1.583pt}{0.111pt}}
\multiput(342.00,479.92)(4.714,6.000){2}{\rule{0.792pt}{1.000pt}}
\multiput(350.00,489.86)(0.151,0.424){2}{\rule{1.650pt}{0.102pt}}
\multiput(350.00,485.92)(3.575,5.000){2}{\rule{0.825pt}{1.000pt}}
\multiput(357.00,494.86)(0.320,0.424){2}{\rule{1.850pt}{0.102pt}}
\multiput(357.00,490.92)(4.160,5.000){2}{\rule{0.925pt}{1.000pt}}
\put(365,497.42){\rule{1.927pt}{1.000pt}}
\multiput(365.00,495.92)(4.000,3.000){2}{\rule{0.964pt}{1.000pt}}
\put(373,500.92){\rule{1.686pt}{1.000pt}}
\multiput(373.00,498.92)(3.500,4.000){2}{\rule{0.843pt}{1.000pt}}
\put(380,504.42){\rule{1.927pt}{1.000pt}}
\multiput(380.00,502.92)(4.000,3.000){2}{\rule{0.964pt}{1.000pt}}
\put(388,506.92){\rule{1.927pt}{1.000pt}}
\multiput(388.00,505.92)(4.000,2.000){2}{\rule{0.964pt}{1.000pt}}
\put(396,509.42){\rule{1.686pt}{1.000pt}}
\multiput(396.00,507.92)(3.500,3.000){2}{\rule{0.843pt}{1.000pt}}
\put(403,511.92){\rule{1.927pt}{1.000pt}}
\multiput(403.00,510.92)(4.000,2.000){2}{\rule{0.964pt}{1.000pt}}
\put(411,513.92){\rule{1.686pt}{1.000pt}}
\multiput(411.00,512.92)(3.500,2.000){2}{\rule{0.843pt}{1.000pt}}
\put(418,515.42){\rule{1.927pt}{1.000pt}}
\multiput(418.00,514.92)(4.000,1.000){2}{\rule{0.964pt}{1.000pt}}
\put(426,516.92){\rule{1.927pt}{1.000pt}}
\multiput(426.00,515.92)(4.000,2.000){2}{\rule{0.964pt}{1.000pt}}
\put(434,518.42){\rule{1.686pt}{1.000pt}}
\multiput(434.00,517.92)(3.500,1.000){2}{\rule{0.843pt}{1.000pt}}
\put(441,519.42){\rule{1.927pt}{1.000pt}}
\multiput(441.00,518.92)(4.000,1.000){2}{\rule{0.964pt}{1.000pt}}
\put(449,520.42){\rule{1.927pt}{1.000pt}}
\multiput(449.00,519.92)(4.000,1.000){2}{\rule{0.964pt}{1.000pt}}
\put(457,521.42){\rule{1.686pt}{1.000pt}}
\multiput(457.00,520.92)(3.500,1.000){2}{\rule{0.843pt}{1.000pt}}
\put(464,522.42){\rule{1.927pt}{1.000pt}}
\multiput(464.00,521.92)(4.000,1.000){2}{\rule{0.964pt}{1.000pt}}
\put(472,523.42){\rule{1.927pt}{1.000pt}}
\multiput(472.00,522.92)(4.000,1.000){2}{\rule{0.964pt}{1.000pt}}
\put(480,524.42){\rule{1.686pt}{1.000pt}}
\multiput(480.00,523.92)(3.500,1.000){2}{\rule{0.843pt}{1.000pt}}
\put(487,525.42){\rule{1.927pt}{1.000pt}}
\multiput(487.00,524.92)(4.000,1.000){2}{\rule{0.964pt}{1.000pt}}
\put(502,526.42){\rule{1.927pt}{1.000pt}}
\multiput(502.00,525.92)(4.000,1.000){2}{\rule{0.964pt}{1.000pt}}
\put(495.0,528.0){\rule[-0.500pt]{1.686pt}{1.000pt}}
\put(518,527.42){\rule{1.686pt}{1.000pt}}
\multiput(518.00,526.92)(3.500,1.000){2}{\rule{0.843pt}{1.000pt}}
\put(510.0,529.0){\rule[-0.500pt]{1.927pt}{1.000pt}}
\put(533,528.42){\rule{1.927pt}{1.000pt}}
\multiput(533.00,527.92)(4.000,1.000){2}{\rule{0.964pt}{1.000pt}}
\put(525.0,530.0){\rule[-0.500pt]{1.927pt}{1.000pt}}
\put(548,529.42){\rule{1.927pt}{1.000pt}}
\multiput(548.00,528.92)(4.000,1.000){2}{\rule{0.964pt}{1.000pt}}
\put(541.0,531.0){\rule[-0.500pt]{1.686pt}{1.000pt}}
\put(571,530.42){\rule{1.927pt}{1.000pt}}
\multiput(571.00,529.92)(4.000,1.000){2}{\rule{0.964pt}{1.000pt}}
\put(556.0,532.0){\rule[-0.500pt]{3.613pt}{1.000pt}}
\put(594,531.42){\rule{1.927pt}{1.000pt}}
\multiput(594.00,530.92)(4.000,1.000){2}{\rule{0.964pt}{1.000pt}}
\put(579.0,533.0){\rule[-0.500pt]{3.613pt}{1.000pt}}
\put(632,532.42){\rule{1.927pt}{1.000pt}}
\multiput(632.00,531.92)(4.000,1.000){2}{\rule{0.964pt}{1.000pt}}
\put(602.0,534.0){\rule[-0.500pt]{7.227pt}{1.000pt}}
\put(678,533.42){\rule{1.927pt}{1.000pt}}
\multiput(678.00,532.92)(4.000,1.000){2}{\rule{0.964pt}{1.000pt}}
\put(640.0,535.0){\rule[-0.500pt]{9.154pt}{1.000pt}}
\put(739,534.42){\rule{1.927pt}{1.000pt}}
\multiput(739.00,533.92)(4.000,1.000){2}{\rule{0.964pt}{1.000pt}}
\put(686.0,536.0){\rule[-0.500pt]{12.768pt}{1.000pt}}
\put(831,535.42){\rule{1.686pt}{1.000pt}}
\multiput(831.00,534.92)(3.500,1.000){2}{\rule{0.843pt}{1.000pt}}
\put(747.0,537.0){\rule[-0.500pt]{20.236pt}{1.000pt}}
\put(1014,536.42){\rule{1.927pt}{1.000pt}}
\multiput(1014.00,535.92)(4.000,1.000){2}{\rule{0.964pt}{1.000pt}}
\put(838.0,538.0){\rule[-0.500pt]{42.398pt}{1.000pt}}
\put(1022.0,539.0){\rule[-0.500pt]{27.463pt}{1.000pt}}
\sbox{\plotpoint}{\rule[-0.175pt]{0.350pt}{0.350pt}}%
\put(251,255){\usebox{\plotpoint}}
\multiput(251.47,255.00)(0.504,8.918){11}{\rule{0.121pt}{5.738pt}}
\multiput(250.27,255.00)(7.000,101.092){2}{\rule{0.350pt}{2.869pt}}
\multiput(258.47,368.00)(0.504,6.585){13}{\rule{0.121pt}{4.331pt}}
\multiput(257.27,368.00)(8.000,88.010){2}{\rule{0.350pt}{2.166pt}}
\multiput(266.47,465.00)(0.504,6.303){11}{\rule{0.121pt}{4.088pt}}
\multiput(265.27,465.00)(7.000,71.516){2}{\rule{0.350pt}{2.044pt}}
\multiput(273.47,545.00)(0.504,4.471){13}{\rule{0.121pt}{2.975pt}}
\multiput(272.27,545.00)(8.000,59.825){2}{\rule{0.350pt}{1.487pt}}
\multiput(281.47,611.00)(0.504,3.652){13}{\rule{0.121pt}{2.450pt}}
\multiput(280.27,611.00)(8.000,48.915){2}{\rule{0.350pt}{1.225pt}}
\multiput(289.47,665.00)(0.504,3.371){11}{\rule{0.121pt}{2.237pt}}
\multiput(288.27,665.00)(7.000,38.356){2}{\rule{0.350pt}{1.119pt}}
\multiput(296.47,708.00)(0.504,2.425){13}{\rule{0.121pt}{1.663pt}}
\multiput(295.27,708.00)(8.000,32.549){2}{\rule{0.350pt}{0.831pt}}
\multiput(304.47,744.00)(0.504,1.948){13}{\rule{0.121pt}{1.356pt}}
\multiput(303.27,744.00)(8.000,26.185){2}{\rule{0.350pt}{0.678pt}}
\multiput(312.47,773.00)(0.504,1.866){11}{\rule{0.121pt}{1.288pt}}
\multiput(311.27,773.00)(7.000,21.328){2}{\rule{0.350pt}{0.644pt}}
\multiput(319.47,797.00)(0.504,1.334){13}{\rule{0.121pt}{0.962pt}}
\multiput(318.27,797.00)(8.000,18.002){2}{\rule{0.350pt}{0.481pt}}
\multiput(327.47,817.00)(0.504,1.129){13}{\rule{0.121pt}{0.831pt}}
\multiput(326.27,817.00)(8.000,15.275){2}{\rule{0.350pt}{0.416pt}}
\multiput(335.47,834.00)(0.504,1.073){11}{\rule{0.121pt}{0.787pt}}
\multiput(334.27,834.00)(7.000,12.366){2}{\rule{0.350pt}{0.394pt}}
\multiput(342.47,848.00)(0.504,0.856){13}{\rule{0.121pt}{0.656pt}}
\multiput(341.27,848.00)(8.000,11.638){2}{\rule{0.350pt}{0.328pt}}
\multiput(350.47,861.00)(0.504,0.756){11}{\rule{0.121pt}{0.587pt}}
\multiput(349.27,861.00)(7.000,8.781){2}{\rule{0.350pt}{0.294pt}}
\multiput(357.47,871.00)(0.504,0.584){13}{\rule{0.121pt}{0.481pt}}
\multiput(356.27,871.00)(8.000,8.001){2}{\rule{0.350pt}{0.241pt}}
\multiput(365.00,880.47)(0.515,0.504){13}{\rule{0.438pt}{0.121pt}}
\multiput(365.00,879.27)(7.092,8.000){2}{\rule{0.219pt}{0.350pt}}
\multiput(373.00,888.47)(0.518,0.504){11}{\rule{0.438pt}{0.121pt}}
\multiput(373.00,887.27)(6.092,7.000){2}{\rule{0.219pt}{0.350pt}}
\multiput(380.00,895.47)(0.712,0.505){9}{\rule{0.554pt}{0.122pt}}
\multiput(380.00,894.27)(6.850,6.000){2}{\rule{0.277pt}{0.350pt}}
\multiput(388.00,901.47)(0.885,0.507){7}{\rule{0.648pt}{0.122pt}}
\multiput(388.00,900.27)(6.656,5.000){2}{\rule{0.324pt}{0.350pt}}
\multiput(396.00,906.47)(0.767,0.507){7}{\rule{0.577pt}{0.122pt}}
\multiput(396.00,905.27)(5.801,5.000){2}{\rule{0.289pt}{0.350pt}}
\multiput(403.00,911.47)(1.183,0.509){5}{\rule{0.787pt}{0.123pt}}
\multiput(403.00,910.27)(6.366,4.000){2}{\rule{0.394pt}{0.350pt}}
\multiput(411.00,915.47)(1.024,0.509){5}{\rule{0.700pt}{0.123pt}}
\multiput(411.00,914.27)(5.547,4.000){2}{\rule{0.350pt}{0.350pt}}
\multiput(418.00,919.47)(1.881,0.516){3}{\rule{1.021pt}{0.124pt}}
\multiput(418.00,918.27)(5.881,3.000){2}{\rule{0.510pt}{0.350pt}}
\multiput(426.00,922.47)(1.881,0.516){3}{\rule{1.021pt}{0.124pt}}
\multiput(426.00,921.27)(5.881,3.000){2}{\rule{0.510pt}{0.350pt}}
\multiput(434.00,925.47)(1.623,0.516){3}{\rule{0.904pt}{0.124pt}}
\multiput(434.00,924.27)(5.123,3.000){2}{\rule{0.452pt}{0.350pt}}
\put(441,928.27){\rule{1.487pt}{0.350pt}}
\multiput(441.00,927.27)(4.913,2.000){2}{\rule{0.744pt}{0.350pt}}
\multiput(449.00,930.47)(1.881,0.516){3}{\rule{1.021pt}{0.124pt}}
\multiput(449.00,929.27)(5.881,3.000){2}{\rule{0.510pt}{0.350pt}}
\put(457,933.27){\rule{1.312pt}{0.350pt}}
\multiput(457.00,932.27)(4.276,2.000){2}{\rule{0.656pt}{0.350pt}}
\put(464,935.27){\rule{1.487pt}{0.350pt}}
\multiput(464.00,934.27)(4.913,2.000){2}{\rule{0.744pt}{0.350pt}}
\put(472,936.77){\rule{1.927pt}{0.350pt}}
\multiput(472.00,936.27)(4.000,1.000){2}{\rule{0.964pt}{0.350pt}}
\put(480,938.27){\rule{1.312pt}{0.350pt}}
\multiput(480.00,937.27)(4.276,2.000){2}{\rule{0.656pt}{0.350pt}}
\put(487,939.77){\rule{1.927pt}{0.350pt}}
\multiput(487.00,939.27)(4.000,1.000){2}{\rule{0.964pt}{0.350pt}}
\put(495,941.27){\rule{1.312pt}{0.350pt}}
\multiput(495.00,940.27)(4.276,2.000){2}{\rule{0.656pt}{0.350pt}}
\put(502,942.77){\rule{1.927pt}{0.350pt}}
\multiput(502.00,942.27)(4.000,1.000){2}{\rule{0.964pt}{0.350pt}}
\put(510,943.77){\rule{1.927pt}{0.350pt}}
\multiput(510.00,943.27)(4.000,1.000){2}{\rule{0.964pt}{0.350pt}}
\put(518,944.77){\rule{1.686pt}{0.350pt}}
\multiput(518.00,944.27)(3.500,1.000){2}{\rule{0.843pt}{0.350pt}}
\put(525,945.77){\rule{1.927pt}{0.350pt}}
\multiput(525.00,945.27)(4.000,1.000){2}{\rule{0.964pt}{0.350pt}}
\put(533,946.77){\rule{1.927pt}{0.350pt}}
\multiput(533.00,946.27)(4.000,1.000){2}{\rule{0.964pt}{0.350pt}}
\put(541,947.77){\rule{1.686pt}{0.350pt}}
\multiput(541.00,947.27)(3.500,1.000){2}{\rule{0.843pt}{0.350pt}}
\put(548,948.77){\rule{1.927pt}{0.350pt}}
\multiput(548.00,948.27)(4.000,1.000){2}{\rule{0.964pt}{0.350pt}}
\put(564,949.77){\rule{1.686pt}{0.350pt}}
\multiput(564.00,949.27)(3.500,1.000){2}{\rule{0.843pt}{0.350pt}}
\put(571,950.77){\rule{1.927pt}{0.350pt}}
\multiput(571.00,950.27)(4.000,1.000){2}{\rule{0.964pt}{0.350pt}}
\put(556.0,950.0){\rule[-0.175pt]{1.927pt}{0.350pt}}
\put(586,951.77){\rule{1.927pt}{0.350pt}}
\multiput(586.00,951.27)(4.000,1.000){2}{\rule{0.964pt}{0.350pt}}
\put(594,952.77){\rule{1.927pt}{0.350pt}}
\multiput(594.00,952.27)(4.000,1.000){2}{\rule{0.964pt}{0.350pt}}
\put(579.0,952.0){\rule[-0.175pt]{1.686pt}{0.350pt}}
\put(609,953.77){\rule{1.927pt}{0.350pt}}
\multiput(609.00,953.27)(4.000,1.000){2}{\rule{0.964pt}{0.350pt}}
\put(602.0,954.0){\rule[-0.175pt]{1.686pt}{0.350pt}}
\put(632,954.77){\rule{1.927pt}{0.350pt}}
\multiput(632.00,954.27)(4.000,1.000){2}{\rule{0.964pt}{0.350pt}}
\put(617.0,955.0){\rule[-0.175pt]{3.613pt}{0.350pt}}
\put(647,955.77){\rule{1.927pt}{0.350pt}}
\multiput(647.00,955.27)(4.000,1.000){2}{\rule{0.964pt}{0.350pt}}
\put(640.0,956.0){\rule[-0.175pt]{1.686pt}{0.350pt}}
\put(670,956.77){\rule{1.927pt}{0.350pt}}
\multiput(670.00,956.27)(4.000,1.000){2}{\rule{0.964pt}{0.350pt}}
\put(655.0,957.0){\rule[-0.175pt]{3.613pt}{0.350pt}}
\put(701,957.77){\rule{1.927pt}{0.350pt}}
\multiput(701.00,957.27)(4.000,1.000){2}{\rule{0.964pt}{0.350pt}}
\put(678.0,958.0){\rule[-0.175pt]{5.541pt}{0.350pt}}
\put(731,958.77){\rule{1.927pt}{0.350pt}}
\multiput(731.00,958.27)(4.000,1.000){2}{\rule{0.964pt}{0.350pt}}
\put(709.0,959.0){\rule[-0.175pt]{5.300pt}{0.350pt}}
\put(770,959.77){\rule{1.686pt}{0.350pt}}
\multiput(770.00,959.27)(3.500,1.000){2}{\rule{0.843pt}{0.350pt}}
\put(739.0,960.0){\rule[-0.175pt]{7.468pt}{0.350pt}}
\put(815,960.77){\rule{1.927pt}{0.350pt}}
\multiput(815.00,960.27)(4.000,1.000){2}{\rule{0.964pt}{0.350pt}}
\put(777.0,961.0){\rule[-0.175pt]{9.154pt}{0.350pt}}
\put(876,961.77){\rule{1.927pt}{0.350pt}}
\multiput(876.00,961.27)(4.000,1.000){2}{\rule{0.964pt}{0.350pt}}
\put(823.0,962.0){\rule[-0.175pt]{12.768pt}{0.350pt}}
\put(968,962.77){\rule{1.927pt}{0.350pt}}
\multiput(968.00,962.27)(4.000,1.000){2}{\rule{0.964pt}{0.350pt}}
\put(884.0,963.0){\rule[-0.175pt]{20.236pt}{0.350pt}}
\put(1105,963.77){\rule{1.927pt}{0.350pt}}
\multiput(1105.00,963.27)(4.000,1.000){2}{\rule{0.964pt}{0.350pt}}
\put(976.0,964.0){\rule[-0.175pt]{31.076pt}{0.350pt}}
\put(1113.0,965.0){\rule[-0.175pt]{5.541pt}{0.350pt}}
\end{picture}
\vspace{0.4cm}
\caption{Illustrative plot for the $CP$-violating signal  
$\xi_{2K}$ changing with $x_s \in [1, 30]$, where $(\phi_m - \phi_K) =
\pi/6$ or $\pi/4$ has been taken.}
\end{figure}
%%%%%%%%%%%%%%%%%%%%%%%%%%%%%%%%%%%

\section{Concluding remarks}

We have studied some phenomenological possibilities to isolate the NP 
effect in $B^0_s$-$\bar{B}^0_s$ mixing. If $x^{\rm SM}_s$ (or $y^{\rm SM}_s$)
can be reliably evaluated and $x_s$ (or $y_s$) can be experimentally 
determined, then significant deviation of $x_s/x^{\rm SM}_s$ (or $y_s/y^{\rm SM}_s$)
from unity will be a clear signal of NP. In addition, NP in $B^0_s$-$\bar{B}^0_s$
mixing is likely to enhance the semileptonic $CP$ asymmetry ${\cal A}_{\rm SL}$
to the level of $10^{-3} - 10^{-2}$. The phase information of NP can be extracted
from either flavor-tagged or flavor-untagged data samples of $B_s$ decays
into hadronic $CP$ eigenstates, or both of them, depending upon the
experimental sensitivity to the $x_s$- and (or) $y_s$-induced oscillations.
It is remarkable that the rates of $CP$-forbidden $B^0_s\bar{B}^0_s$ transitions
on the $\Upsilon (5S)$ resonance are not suppressed by the largeness of
$x_s$, and this feature will be useful for measuring the $CP$-violating phase in 
$B^0_s$-$\bar{B}^0_s$ (or $K^0$-$\bar{K}^0$) mixing.

Of course, the experimental feasibility of the above-proposed measurements
has to be studied in some detail \cite{Xing98}. In particular, the complication 
to observe coherent $B^0_s\bar{B}^0_s$ decays at the $\Upsilon (5S)$
resonance, where coherent $B^0_d\bar{B}^0_d$ events can also be produced 
\cite{Xing98,Krawczyk}, should be taken seriously. 
We hope that a systematic search for various $CP$-violating signals in the
$B^0_s$-$\bar{B}^0_s$ system will be available in the future experiments
of $B$-meson physics.

\vspace{0.3cm}

\begin{flushleft}
{\Large\bf Acknowledgments}
\end{flushleft}

The author likes to thank A.I. Sanda for some useful discussions.
This work was supported by the Japan Society for the Promotion of
Science.

\end{document}